\documentclass[12pt]{article}

\usepackage{amsmath,amsthm,amssymb} 
\usepackage{mathbbol} 
\usepackage{enumitem} 
\usepackage{graphicx} 
\usepackage[T1]{fontenc} 
\usepackage[numbered,framed]{matlab-prettifier}
\usepackage{xcolor} 

\usepackage{hyperref} 

\usepackage[labelformat=simple]{subcaption} 

\theoremstyle{definition}

\title{\textsc{LogLatt}: A computational library for the \\calculus and flows on logarithmic lattices}
\author{Ciro S. Campolina\thanks{Instituto de Matem\'{a}tica Pura e Aplicada -- IMPA, 22460-320 Rio de Janeiro, Brazil. E-mail: \texttt{sobrinho@impa.br.}}}

\date{}

\begin{document}
	
	\maketitle
	
	\begin{abstract}
		Models on logarithmic lattices have recently been proposed as an alternative approach to the study of multi-scale nonlinear physics.
		Here, we introduce \textsc{LogLatt}, an efficient \textsc{Matlab}\textsuperscript{\circledR} library for the calculus between functions on multi-dimensional logarithmic lattices.
		The applicabilities include common differential operators, norms, and convolutions,
		and operate as \textsc{Matlab} function handles, so their implementations result in elegant and intuitive scripts.
		Particularly, when applied to dynamics, users may code the governing equations exactly as they are mathematically written.
		We give codes and results for some problems of fluid flow as examples of application.
	\end{abstract}
	
	\section{Introduction}
	
	Many problems in nonlinear science are related to the spontaneous development of small-scale structures.
	This is the scenario for several open questions in fluid dynamics, such as the global regularity of the Navier-Stokes equations~\cite{fefferman2006existence}, the formation of singularities in ideal flow~\cite{gibbon2008three}, and the statistical description of intermittent turbulence~\cite{frisch1995turbulence}.
	In most cases, their multi-scale nature is inaccessible to the resolution of current computational techniques.
	This fact attests the difficulty in solving such problems, and reveals the need of alternative methods of investigation.
	
	To overcome the above limitations, a new technique was recently proposed~\cite{campolina2018chaotic}.
	It consists of simulating the governing equations on multi-dimensional lattices of logarithmically distributed nodes in Fourier space.
	A special calculus was designed for this domain~\cite{campolina2021fluid}.
	As a result, the equations of motion preserve their exact form and retain most of the properties of the original systems, like the symmetry groups and conserved quantities.
	Their strong reduction in degrees of freedom allows the simplified models to be easily simulated on a computer within a surprisingly large spatial range.
	This method was successfully applied to important problems in fluid dynamics, such as the blowup and shock solutions in the Burgers equation~\cite{campolina2019fluid}, the chaotic blowup in ideal flow~\cite{campolina2018chaotic}, and the Navier-Stokes turbulence~\cite{campolina2021fluid}.
	More generally, this technique is ready-to-use on any differential equation with quadratic nonlinearity.
	
	However, some special operations on logarithmic lattices may result in nontrivial computational implementations.
	As an example, the main operation on the lattice is the product between functions, which becomes a convolution in Fourier space.
	The geometry of the lattice turns this product into an unconventional discrete convolution coupling specific local triads.
	Therefore, its numerical computation requires a prior complicated classification of interacting triads on the lattice.
	
	Here, we introduce \textsc{LogLatt}, an efficient \textsc{Matlab}\textsuperscript{\circledR} library for the numerical calculus and operations between functions on logarithmic lattices.
	It is freely available for noncommercial use in \textsc{Matlab} Central File Exchange~\cite{campolina2020loglattmatlab}.
	The computational applicabilities are available for one-, two- and three-dimensional lattices, and include usual differential operators from vector calculus, norms and the aforementioned functional products.
	The operations are encoded as \textsc{Matlab} function handles. This provides simple and intuitive scripts.
	When applied to partial differential equations, the models are coded exactly as they are mathematically written.
	This library will make the computational calculus on logarithmic lattices accessible in an efficient framework for the study of nonlinear equations.
	
	The paper is divided as follows.
	In Section~\ref{SEC:calculus_log} we briefly review the calculus on logarithmic lattices developed in~\cite{campolina2021fluid}.
	We describe the library, its implementation and computational cost in Section~\ref{SEC:library}, and apply it to some problems of fluid flow in Section~\ref{SEC:fluid}.
	We address the conclusions in Section~\ref{SEC:conclusions}.
	
	\section{Calculus on logarithmic lattices}\label{SEC:calculus_log}
	
\begin{figure*}[t]
		\centering
		\includegraphics[width=\textwidth]{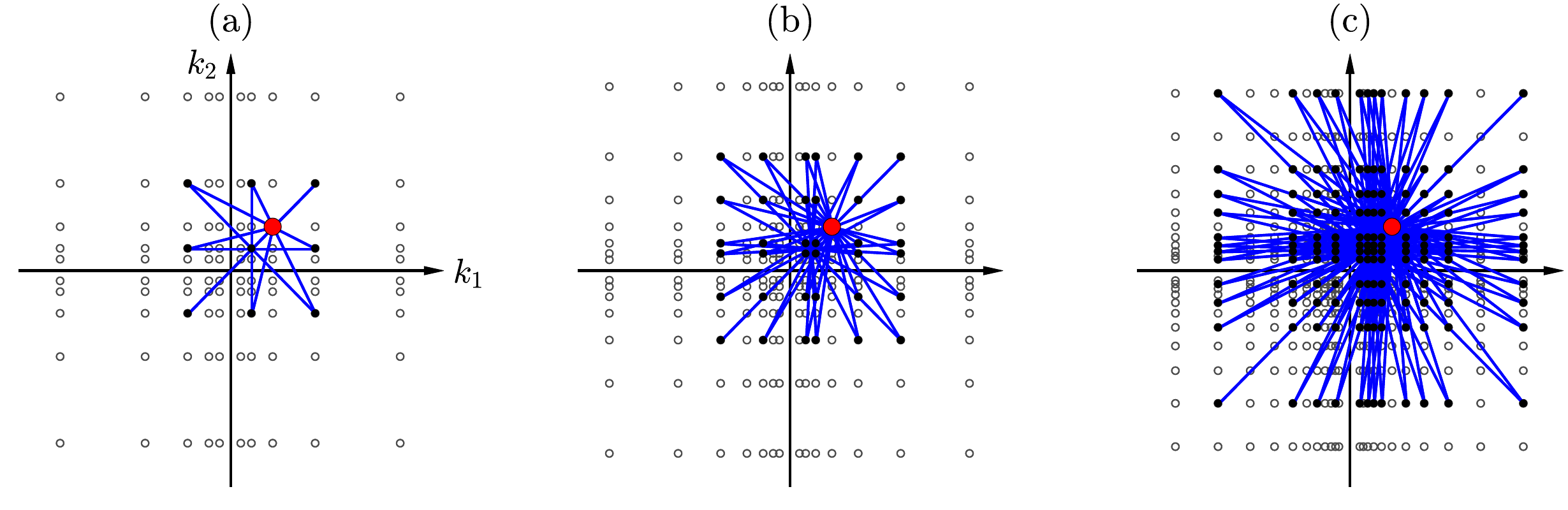}
		\caption{Triad interactions on two-dimensional logarithmic lattices of different spacings: (a)~$\lambda = 2$; (b)~$\lambda = \varphi \approx 1.618$, the golden mean; (c)~$\lambda = \sigma \approx 1.325$, the plastic number. The red node $\mathbf{k}$ can be decomposed into sums $\mathbf{k} = \mathbf{p} + \mathbf{q}$ where all possible nodes $\mathbf{p}$ and $\mathbf{q}$ are indicated by the blue lines. All figures are given in the same scale. From (a) to (c), both the density of nodes and the number of triads per each node increase.}
		\label{FIG:2D_lattice}
	\end{figure*}

	We start with the definitions of logarithmic lattices.
	Given a number $\lambda > 1$, the \textit{one-dimensional logarithmic lattice of spacing} $\lambda$ is the set
	\begin{equation}
	\mathbb{\Lambda} = \{ \pm \lambda^n \}_{n \in \mathbb{Z}},
	\label{EQ:log_lattice}
	\end{equation}
	consisting of positive and negative integer powers of $\lambda$.
	This set is scale invariant, \textit{i.e.}, $\mathbb{\Lambda} = k \mathbb{\Lambda}$ for every $k \in \mathbb{\Lambda}$.
	In practice, we truncate the infinite sequence~\eqref{EQ:log_lattice} and employ only a finite number of points
	\begin{equation}
	\mathbb{\Lambda} = \{ \pm \lambda^{-(M-1)}, \dots, \pm \lambda^{-2}, \pm \lambda^{-1}, \pm 1, \pm \lambda, \pm \lambda^2, \dots, \pm \lambda^{N-1} \},
	\label{EQ:truncated_lattice}
	\end{equation}
	which recover the original set by considering the limit $M,N \to \infty$.
	All operations and properties presented below equally apply to the infinite~\eqref{EQ:log_lattice} or the truncated~\eqref{EQ:truncated_lattice} lattices.
	The \textit{$d$-dimensional logarithmic lattice of spacing} $\lambda$ is given by the Cartesian power $\mathbb{\Lambda}^d$.
	In this case, $\mathbf{k} = (k_1,\dots,k_d) \in \mathbb{\Lambda}^d$ if each component $k_j \in \mathbb{\Lambda}$.
	See Fig.~\ref{FIG:2D_lattice} for examples of two-dimensional lattices.
	
	We consider complex valued functions $f(\mathbf{k}) \in \mathbb{C}$ on the logarithmic lattice, with $\mathbf{k} \in \mathbb{\Lambda}^d$ interpreted as a wave vector in Fourier space.
	Motivated by the property of the Fourier transform of a real-valued function, we impose the reality condition
	\begin{equation}
	f(-\mathbf{k}) = \overline{f(\mathbf{k})},
	\label{EQ:reality}
	\end{equation}
	where the overline is the complex conjugation.
	Therefore, functions $f(\mathbf{k})$ mimic the Fourier transform of real-valued functions and possess a natural structure of linear space upon real scalars.
	Given two functions $f$ and $g$, their \textit{inner product} is
	\begin{equation}
	(f,g) = \sum_{\mathbf{k} \in \mathbb{\Lambda}^d}f(\mathbf{k}) \overline{g(\mathbf{k})},
	\label{EQ:inner_product}
	\end{equation}
	which is real valued because of reality condition~\eqref{EQ:reality}, and the induced $\ell^2$ norm is
	\begin{equation}
	\Vert f \Vert = (f,f)^{1/2} = \left(\sum_{\mathbf{k} \in \mathbb{\Lambda}^d}|f(\mathbf{k})|^2\right)^{1/2}.
	\label{EQ:l2norm}
	\end{equation}
	
	Differentiation on the lattice is introduced as follows.
	The \textit{partial derivative $\partial_j$ in the $j$-th direction} comes from the Fourier factor
	\begin{equation}
	\partial_j f(\mathbf{k}) = ik_j f(\mathbf{k}),
	\label{EQ:derivative}
	\end{equation}
	where $i$ is the imaginary unit.
	Common differential operators are naturally obtained from definition~\eqref{EQ:derivative}, \textit{e.g.} the gradient $\text{grad} \ f (\mathbf{k})= (\partial_1f, \dots, \partial_d f) = i(k_1f,\dots,k_df)$ and the laplacian $\Delta f (\mathbf{k})= (\partial^2_1 + \cdots + \partial^2_d)f = -|\mathbf{k}|^2f$.
	The divergence $\text{div} \ \mathbf{u}$ and the rotational $\text{rot} \ \mathbf{u}$ are defined in the same manner for vector fields $\mathbf{u}(\mathbf{k}) \in \mathbb{C}^d$ on the lattice $\mathbf{k} \in \mathbb{\Lambda}^d$.
	This notion of differentiability on the lattice retains most properties of calculus.
	Some examples are many vector calculus identities, such as $\text{div} \ \text{grad} = \Delta$, $\text{div} \ \text{rot} = 0$ and $\text{rot} \ \text{grad} = \mathbf{0}$, and integration by parts, $(\partial_j f, g) = - (f,\partial_j g)$ for $j = 1,\dots,d$.
	
	Next, we introduce the product of functions as a discrete convolution on the logarithmic lattice.
	Given two functions $f$ and $g$, their \textit{product} $f \ast g$ is the function
	\begin{equation}
	(f \ast g)(\mathbf{k}) = \sum_{\substack{\mathbf{p} + \mathbf{q} = \mathbf{k}\\[2pt] \mathbf{p},\mathbf{q} \in \mathbb{\Lambda}^d}} f(\mathbf{p}) g(\mathbf{q}), \quad \mathbf{k} \in \mathbb{\Lambda}^d,
	\label{EQ:convolution}
	\end{equation}
	coupling triads $\mathbf{k} = \mathbf{p} + \mathbf{q}$ on the lattice $\mathbf{k},\mathbf{p},\mathbf{q} \in \mathbb{\Lambda}^d$.
	This operation satisfies the following properties:
	\begin{enumerate}[label=\textit{(P.\arabic*)}] 
		\item \emph{(reality condition)}
		$
		\!
		\begin{aligned}[t]
		(f \ast g)(-\mathbf{k}) = \overline{(f \ast g) (\mathbf{k})};
		\end{aligned}
		$ 
		\label{DEF:PROD_reality}
		\item \emph{(bilinearity)} 
		$
		\!
		\begin{aligned}[t]
		(f + \gamma g) \ast h = f \ast h + \gamma(g \ast h),
		\end{aligned}
		$
		for any $\gamma \in \mathbb{R}$;
		\label{DEF:PROD_bilinearity}
		\item \emph{(commutativity)}
		$
		\!
		\begin{aligned}[t]
		f \ast g = g \ast f;
		\end{aligned}
		$ 
		\label{DEF:PROD_commutativity}
		\item \emph{(associativity in average)}
		$
		\!
		\begin{aligned}[t]
		(f \ast g, h ) = (f, g \ast h),
		\end{aligned}
		$
		with respect to inner product~\eqref{EQ:inner_product};
		\label{DEF:PROD_associativity_avg}
		\item \emph{(translation invariance)}
		$
		\!
		\begin{aligned}[t]
		\tau_{\pmb{\xi}} (f \ast g) = \tau_{\pmb{\xi}} f \ast \tau_{\pmb{\xi}} g,
		\end{aligned}
		$
		where $\tau_{\pmb{\xi}} f (\pmb{k}) = e^{-i\pmb{k} \cdot \pmb{\xi}} f(\pmb{k})$ mimics the physical-space translation (in Fourier representation) by any vector $\pmb{\xi} \in \mathbb{R}^d$;
		\label{DEF:PROD_translation}
		\item \emph{(isotropy and parity)} 
		$
		\!
		\begin{aligned}[t]
		(f\ast g) \circ R = (f \circ R) \ast (g \circ R),
		\end{aligned}
		$
		where $(f \circ R)(\mathbf{k}) = f(R\mathbf{k})$, and $R \in \mathsf{O_h}$ is any element of the group of cube symmetries. This group includes all transformations $(k_1,\dots,k_d) \mapsto (\pm k_{\alpha_1},\dots,\pm k_{\alpha_d})$, where $(\alpha_1,\dots,\alpha_d)$ are permutations of $(1,\dots,d)$ -- \textit{cf.}~\cite[Sec.~93]{landau2013quantum};
		\label{DEF:PROD_isotropy}
		\item \emph{(Leibniz rule)}
		$
		\!
		\begin{aligned}[t]
		\partial_j (f \ast g) = \partial_j f \ast g + f \ast \partial_j g, \ \text{for } j = 1,\dots,d.
		\end{aligned}
		$
		\label{DEF:PROD_leibniz}
	\end{enumerate}
	On the infinite lattice~\eqref{EQ:log_lattice}, product~\eqref{EQ:convolution} also satisfies
	\begin{enumerate}[label=\textit{(P.8)}] 
		\item \emph{(scaling invariance)}
		$
		\!
		\begin{aligned}[t]
		\delta_{\lambda}(f \ast g) = \delta_{\lambda} f \ast \delta_{\lambda} g,
		\end{aligned}
		$
		where we denoted the rescaling of $f$ by the lattice spacing $\lambda$ as $\delta_\lambda f (\mathbf{k}) = f(\lambda \mathbf{k})$.
		\label{DEF:PROD_scaling}
	\end{enumerate}
	Scaling invariance~\ref{DEF:PROD_scaling} is also satisfied in truncated lattices~\eqref{EQ:truncated_lattice} if we consider zero padding of the functions at the excluded points.
	
	Properties~\ref{DEF:PROD_reality}--\ref{DEF:PROD_scaling} mimic as much as possible the usual properties of a convolution in Fourier space.
	We remark that the operation~\eqref{EQ:convolution} is not associative, \textit{i.e.}, $(f\ast g) \ast h \neq f \ast (g \ast h)$ for general $f,g$ and $h$.
	Nevertheless, associativity in average~\ref{DEF:PROD_associativity_avg} is valid, which is sufficient for many applications in equations with quadratic nonlinearity, such as those from incompressible fluid dynamics.
	
	\begin{table}[t]
		\begin{subtable}{.35\textwidth}
			\caption{} 
			\centering 
			\begin{tabular}{c c c c} 
				\hline\hline 
				$i$&$\phantom{-}1$&$\phantom{-}2$&$3$ \\
				\hline
				$p_i$&$\phantom{-}2$&$-1$&$1/2$ \\ 
				$q_i$&$-1$&$\phantom{-}2$&$1/2$ \\
				\hline\hline
			\end{tabular}
			\label{TAB:DN}
		\end{subtable}\hspace*{.05\textwidth}
		\begin{subtable}{.6\textwidth}
			\caption{} 
			\centering 
			\begin{tabular}{c c c c c c c} 
				\hline\hline 
				$i$&$\phantom{-}1$&$\phantom{-}2$&$3$&$4$&$5$&$6$ \\
				\hline
				$p_i$&$\phantom{-}\lambda^b$&$-\lambda^a$&$\phantom{-}\lambda^{b-a}$&$-\lambda^{-a}$&$\lambda^{-b}$&$\lambda^{a-b}$ \\ 
				$q_i$&$-\lambda^a$&$\phantom{-}\lambda^b$&$-\lambda^{-a}$&$\phantom{-}\lambda^{b-a}$&$\lambda^{a-b}$&$\lambda^{-b}$ \\
				\hline\hline
			\end{tabular}
			\label{TAB:general}
		\end{subtable}\vspace*{20pt}
		\begin{subtable}{\textwidth}
			\caption{} 
			\centering 
			\begin{tabular}{c c c c c c c c c c c c c} 
				\hline\hline 
				$i$&$\phantom{-}1$&$\phantom{-}2$&$3$&$4$&$5$&$6$&$\phantom{-}7$&$\phantom{-}8$&$9$&$10$&$11$&$12$ \\
				\hline
				$p_i$&$\phantom{-}\sigma^3$&$-\sigma$&$\sigma^2$&$-\sigma^{-1}$&$\sigma^{-3}$&$\sigma^{-2}$&
				$\phantom{-}\sigma^5$&$-\sigma^4$&$\phantom{-}\sigma^{\phantom{-}}$&$-\sigma^{-4}$&$\sigma^{-5}$&$\sigma^{-1}$ \\
				$q_i$&$-\sigma^{\phantom{-}}$&$\phantom{-}\sigma^3$&$-\sigma^{-1}$&$\sigma^2$&$\sigma^{-2}$&$\sigma^{-3}$&
				$-\sigma^4$&$\phantom{-}\sigma^5$&$-\sigma^{-4}$&$\sigma$&$\sigma^{-1}$&$\sigma^{-5}$ \\
				\hline\hline
			\end{tabular}
			\label{TAB:extended}
		\end{subtable}
		\caption{Triads at the unity $1 = p_i + q_i$ for different lattice spacings: \subref{TAB:DN}~$\lambda = 2$;
			\subref{TAB:general}~$\lambda$ satisfies $1 = \lambda^b-\lambda^a$ for integers $0 \leq a <b$. For example, $\lambda$ is the golden mean $\varphi$ for $a = 1$ and $b = 2$;
			\subref{TAB:extended}~$\lambda = \sigma$, the plastic number.}
		\label{TAB:triads_unity}
	\end{table}
	
	However, the product~\eqref{EQ:convolution} is nontrivial only for certain spacings $\lambda$.
	In fact, logarithmic lattices are not closed under addition as $\mathbf{p} + \mathbf{q} \notin \mathbb{\Lambda}^d$ for general $\mathbf{p},\mathbf{q} \in \mathbb{\Lambda}^d$, so the existence and the calculation of triads $\mathbf{k} = \mathbf{p} + \mathbf{q}$ are subordinated to the choice of $\lambda$.
	For the classification of such lattices, it suffices to classify the triads at the unity $1 = p + q$, with $p,q$ in the one-dimensional lattice $\mathbb{\Lambda}$.
	Indeed, this follows from two facts:
	\textit{(i)} the lattice is scale invariant, so general triads are rescaled from those at the unity;
	\textit{(ii)} triads on $d$-dimensional logarithmic lattices are combinations of one-dimensional triads, component by component.
	The following are three lattices and their triads:
	\begin{enumerate}[label=\textit{(L.\arabic*)}] 
		\item \textit{(dyadic)} $\lambda = 2$, and all triads at the unity are given in Tab.~\ref{TAB:DN};
		\label{L:dyadic}
		\item \textit{(golden mean)} $\lambda = \varphi$, where $\varphi = (1+\sqrt{5})/2 \approx 1.618$ is the golden mean, and all triads at the unity are given in Tab.~\ref{TAB:general} for $a = 1$ and $b=2$;
		\label{L:golden}
		\item \textit{(plastic number)} $\lambda = \sigma$, where $\sigma = (\sqrt[3]{9+\sqrt{69}}+\sqrt[3]{9-\sqrt{69}})/\sqrt[3]{18} \approx 1.325$ is the plastic number, and all triads at the unity are given in Tab.~\ref{TAB:extended}.
		\label{L:plastic}
	\end{enumerate}
	These and other lattices with triads may be obtained also from
	\begin{enumerate}[label=\textit{(L.4)}] 
		\item \textit{(integers $0\leq a < b$)} $\lambda$ satisfies $1 = \lambda^b - \lambda^a$, where $0 \leq a < b$ are some integers, and triads at the unity are given in Tab.~\ref{TAB:general}.
		\label{L:general}
	\end{enumerate}
	See Fig.~\ref{FIG:2D_lattice} for the two-dimensional lattices \ref{L:dyadic}--\ref{L:plastic} and their triads.
	From\ref{L:dyadic} to \ref{L:plastic}, both the density of nodes and the number of triads per each node increase, thus providing finer resolution.
	In effect, it was shown that these are all possible lattices important for applications -- consult~\cite{campolina2021fluid} for the precise statement and its proof.
	
	We say that the lattice is \textit{nondegenerate} if every two nodes interact though a finite sequence of triads.
	In this case, product~\eqref{EQ:convolution} cannot be decomposed into sums of non-interacting nodes, which would decouple the dynamics into isolated subsystems.
	Lattices~\ref{L:dyadic}--\ref{L:plastic} are nondegenerate, while~\ref{L:general} is nondegenerate only if $a$ and $b$ are mutually prime integers.
	
	\section{Computational library}\label{SEC:library}
	
	We turn now to the numerical implementation of operations described in the previous section.
	Since only a finite number of points can be represented on computer's memory, all routines are developed upon the truncated lattice
	\begin{equation}
	\mathbb{\Lambda} = \{ \pm 1, \pm \lambda, \pm \lambda^2, \dots, \pm \lambda^{N-1} \}
	\label{EQ:computational_lattice}
	\end{equation}
	with $N$ points in each direction.
	Then, lattice $\mathbb{\Lambda}^d$ mimics the $d$-dimensional Fourier space of a system with largest integral scale $L \sim 2\pi$ corresponding to $|\mathbf{k}| \sim 1$ and finest scale $\ell$ defined by $N$ as $\ell \sim 2\pi/\lambda^{N-1}$.
	Finer resolutions may be accessed by increasing $N$.
	
	This section is subdivided as follows.
	\textsc{LogLatt} is composed by distinct routines for each of the spatial dimensions.
	We first present the simpler one-dimensional case in Section~\ref{SEC:1D} to next extend it to two and three dimensions in Section~\ref{SEC:ND}.
	Here the focus is to describe the applicabilities of the library and how to implement each operation.
	Lastly, we discuss in Section~\ref{SEC:cost} the computational efficiency of the products, the most expensive operation of the library.
	
	\subsection{One-dimensional lattices}\label{SEC:1D}
	
	In one-dimensional space, the logarithmic lattice is simply the set~\eqref{EQ:computational_lattice}.
	Because of reality condition~\eqref{EQ:reality}, we don't need to carry negative lattice points on the memory.
	Therefore, functions $f(k) \in \mathbb{C}$ on the lattice $k \in \mathbb{\Lambda}$ are represented by complex valued arrays {\mlttfamily f} of size $N \times 1$.
	
	All applicabilities in the one-dimensional case are encoded in the m-file {\mlttfamily LogLatt1D.m}.
	This routine should be called by specifying the lattice, say the number of points $N$ and the lattice spacing $\lambda$.
	This can be done in three different forms of input:
	\begin{enumerate}[label=\textit{(I.\arabic*)}]
		\item {\mlttfamily(N)} If only the number $N$ of nodes is input, the lattice spacing is the golden mean~\ref{L:golden}, by default;
		\label{I_default}
		\item {\mlttfamily(N,str)} The usual lattice spacings~\ref{L:dyadic}, \ref{L:golden} and \ref{L:plastic} may be input as strings {\mlttfamily str~= \textquotesingle dyadic\textquotesingle}, {\mlttfamily \textquotesingle golden\textquotesingle} and {\mlttfamily \textquotesingle plastic\textquotesingle}, respectively;
		\label{I_str}
		\item {\mlttfamily(N,a,b)} Lattice~\ref{L:general} may be introduced through the integers $0 \leq a < b$.
		\label{I_integers}
	\end{enumerate}
	Henceforward, we adopt input~\ref{I_str} whenever the function needs the lattice to be specified.
	The operations and the lattice itself are obtained through the command {\mlttfamily[product,l2norm, l2inner,sup,dx,lapl,K] = LogLatt1D(N,str)}.
	We describe now each of the outputs.
	
	The array {\mlttfamily K} has size $N \times 1$ and contains the lattice points $(1,\lambda,\lambda^2,\dots,\lambda^{N-1})$.
	The remaining outputs are function handles for the operations on lattice functions:
	given two functions $f$ and $g$, encoded as $N \times 1$ arrays {\mlttfamily f} and {\mlttfamily g}, {\mlttfamily l2inner(f,g)} returns their inner product~\eqref{EQ:inner_product};
	{\mlttfamily l2norm(f)} and {\mlttfamily sup(f)} give the $\ell^2$ norm~\eqref{EQ:l2norm} and the maximum absolute value $\max_{k \in \mathbb{\Lambda}}|f(k)|$ of $f$;
	{\mlttfamily dx(f)} computes the function $\partial_x f$, which is the spatial derivative of $f$ given by the Fourier factor~\eqref{EQ:derivative};
	{\mlttfamily lapl(f)} is the laplacian of $f$, which in one-dimensional space is simply the second-order spatial derivative $\partial_x^2 f$;
	finally, {\mlttfamily product(f,g)} gives the product function $f \ast g$, \textit{i.e.}, the convolution on the lattice~\eqref{EQ:convolution}.
	A practical example of implementation will be given in Section~\ref{SEC:Burgers}, where the operations from {\mlttfamily LogLatt1D.m} are applied in the study of the one-dimensional Burgers equation.
	
	\subsection{Two- and three-dimensional lattices}\label{SEC:ND}
	
	We describe in details the two-dimensional case only, which, except for minor changes enumerated at the end of this section, has the same form of implementation and applicabilities for three-dimensional lattices.
	
	\medskip
	\textsc{Two dimensions.} Functions on two-dimensional logarithmic lattices are computationally represented by matrices.
	Because of reality condition~\eqref{EQ:reality}, it is sufficient to keep only the first two quadrants of Fourier space.
	In this case, scalar functions $f$ are encoded as complex matrices {\mlttfamily f} of size $N \times N \times 2$, where
	{\mlttfamily f(m,n,q)} returns the value $f(\mathbf{k}_{m,n,q})$ with
	\begin{equation}
	\mathbf{k}_{m,n,q} =
	\begin{cases}
	(\phantom{-}\lambda^{m-1},\phantom{-}\lambda^{n-1}) \quad &\text{if} \ q=1 \quad \textit{(1st \phantom{-}quadrant)},\\
	(-\lambda^{m-1},\phantom{-}\lambda^{n-1}) \quad &\text{if} \ q=2 \quad \textit{(2nd quadrant)}.
	\end{cases}
	\label{EQ:2D_enumeration}
	\end{equation}
	Vector fields receive an additional input at the end, which designates the component, and thus become of size $N\times N \times 2 \times 2$.
	
	The applicabilities for two-dimensional lattices are split into four routines: {\mlttfamily LogLatt2D.m}, for the geometry of the lattice; {\mlttfamily LogLatt2D\_diff.m}, for the two-dimensional differential operators; {\mlttfamily LogLatt2D\_norms.m}, for the norms; and {\mlttfamily LogLatt2D\_product.m}, for the functional product.
	Except for the norms routine, which receives no inputs, all of them are called by specifying the lattice in one of the three ways~\ref{I_default}--\ref{I_integers} above.
	We describe their outputs in details now.
	
	The lattice itself is obtained by the command {\mlttfamily [Kx,Ky,Knorm] = LogLatt2D(N,str)}: {\mlttfamily Kx} and {\mlttfamily Ky} are the $x$ and $y$ coordinates of each point from the lattice; {\mlttfamily Knorm} is their Euclidean norm.
	They are $N \times N \times 2$ matrices and their entries follow the same enumeration~\eqref{EQ:2D_enumeration} as described above for scalar functions.
	
	For the two-dimensional lattice, several differential operators from vector calculus are available as function handles through the command {\mlttfamily [dx,dy,lapl,lapl\_,grad,div,rot, rot\_] = LogLatt2D\_diff(N,str)}:
	given a scalar function $f$ encoded as the matrix {\mlttfamily f}, {\mlttfamily dx(f)} and {\mlttfamily dy(f)} are the partial derivatives $\partial_xf$ and $\partial_yf$;
	{\mlttfamily lapl(f)} and {\mlttfamily lapl\_(f)} are the laplacian operator $\Delta f(\mathbf{k}) = -|\mathbf{k}|^2f(\mathbf{k})$ and its inverse $\Delta^{-1} f(\mathbf{k}) = -|\mathbf{k}|^{-2}f(\mathbf{k})$, which is well-defined on the lattice~\eqref{EQ:computational_lattice};
	{\mlttfamily grad(f)} computes the gradient $\text{grad} \ f = (\partial_x f, \partial_y f)$;
	given a vector field $\mathbf{u} = (u_x,u_y)$ represented by the matrix {\mlttfamily u}, {\mlttfamily div(u)} calculates its divergence $\text{div} \ \mathbf{u} = \partial_x u_x + \partial_y u_y$ and {\mlttfamily rot(u)} its scalar rotational $\text{rot} \ \mathbf{u} = \partial_x u_y - \partial_y u_x$, \textit{i.e.}, the nontrivial $z$-component of the full rotational vector;
	{\mlttfamily rot\_} is the inverse of rotational in the space of solenoidal vector fields; more precisely, it receives a scalar function $f$ and computes an incompressible vector field $\mathbf{u}$ satisfying $\text{rot} \ \mathbf{u} = f$, explicitly given by $\mathbf{u} = -\Delta^{-1}(\partial_yf, -\partial_x f)$.
	
	The norms are initialized by {\mlttfamily [l2norm,l2inner,sup] = LogLatt2D\_norms}, with no inputs.
	They are implemented as it was described in the one-dimensional case, and they operate equally on scalar functions and vector fields.
	
	Finally, product~\eqref{EQ:convolution} is obtained from {\mlttfamily product = LogLatt2D\_product(N,str)} and operates on scalar functions only.
	
	\medskip
	\textsc{Three dimensions.}
	Here we limit ourselves to highlight the differences between the three- and two-dimensional routines.
	Because of reality condition~\eqref{EQ:reality}, we only represent the four octants with $z>0$ in the three-dimensional space.
	Extending the two-dimensional enumeration~\eqref{EQ:2D_enumeration}, scalar functions $f$ are encoded as complex matrices {\mlttfamily f} of size $N \times N \times N \times 4$, where {\mlttfamily f(m,n,p,q)} returns the value $f(\mathbf{k}_{m,n,p,q})$ with
	\begin{equation}
	\mathbf{k}_{m,n,p,q} =
	\begin{cases}
	(\phantom{-}\lambda^{m-1},\phantom{-}\lambda^{n-1},\phantom{-}\lambda^{p-1}) \quad &\text{if} \ q=1 \quad \textit{(1st \phantom{-}octant)},\\
	(-\lambda^{m-1},\phantom{-}\lambda^{n-1},\phantom{-}\lambda^{p-1}) \quad &\text{if} \ q=2 \quad \textit{(2nd octant)},\\
	(-\lambda^{m-1},-\lambda^{n-1},\phantom{-}\lambda^{p-1}) \quad &\text{if} \ q=3 \quad \textit{(3rd \phantom{-}octant)},\\
	(\phantom{-}\lambda^{m-1},-\lambda^{n-1},\phantom{-}\lambda^{p-1}) \quad &\text{if} \ q=4 \quad \textit{(4th \phantom{-}octant)}.
	\end{cases}
	\end{equation}
	Vector fields receive an additional entry in the end indicating the component, and thus become of size $N \times N \times N \times 4 \times 3$.
	Routines for geometry and differential operators output the additional $z$-component of the lattice {\mlttfamily Kz} and the partial derivative in the $z$-direction {\mlttfamily dz}.
	The rotational {\mlttfamily rot} gives the full vector field $\text{rot} \ \mathbf{u} = (\partial_yu_z - \partial_zu_y, \partial_zu_x - \partial_xu_z, \partial_xu_y - \partial_yu_x)$ and its inverse in the space of solenoidal fields {\mlttfamily rot\_} receives a vector field $\mathbf{u}$ and computes the incompressible vector field $\mathbf{v}$ satisfying $\text{rot} \ \mathbf{v} = \mathbf{u}$, explicitly given by $\mathbf{v} = -\Delta^{-1} \text{rot} \ \mathbf{u}$.
	\bigskip
	
	In Section~\ref{SEC:Euler}, we show how to apply the two-dimensional library to solve the incompressible Euler equations on the two-dimensional lattice.
	
	\subsection{Computational cost}\label{SEC:cost}
	
	Computationally, the most expensive operation on the logarithmic lattice is the unconventional convolution~\eqref{EQ:convolution}, which couples local triads in Fourier space.
	When applied to nonlinear differential equations, the many executions of such products may take a substantial parcel of the computational cost.
	Aiming to reduce the execution time, \textsc{LogLatt} adopts the following coding strategy.
	When initialized, the product routine locates and stores all interacting triads into an array mask, which is efficiently used each time the product is invoked.
	This strategy reduces the time spent on the computation of each convolution, at the cost of, first and only once, classifying and storing in memory all triads in the lattice.
	In two and three dimensions, this is done by an external C routine, called by \textsc{Matlab} through a mex-file, which, for larger amounts of storage, proved to be more efficient than solely the \textsc{Matlab} computation.
	The one-dimesional case, instead, benefits from \textsc{Matlab}'s multithreading.
	
\begin{figure*}[t]
		\centering
		\includegraphics[width=\textwidth]{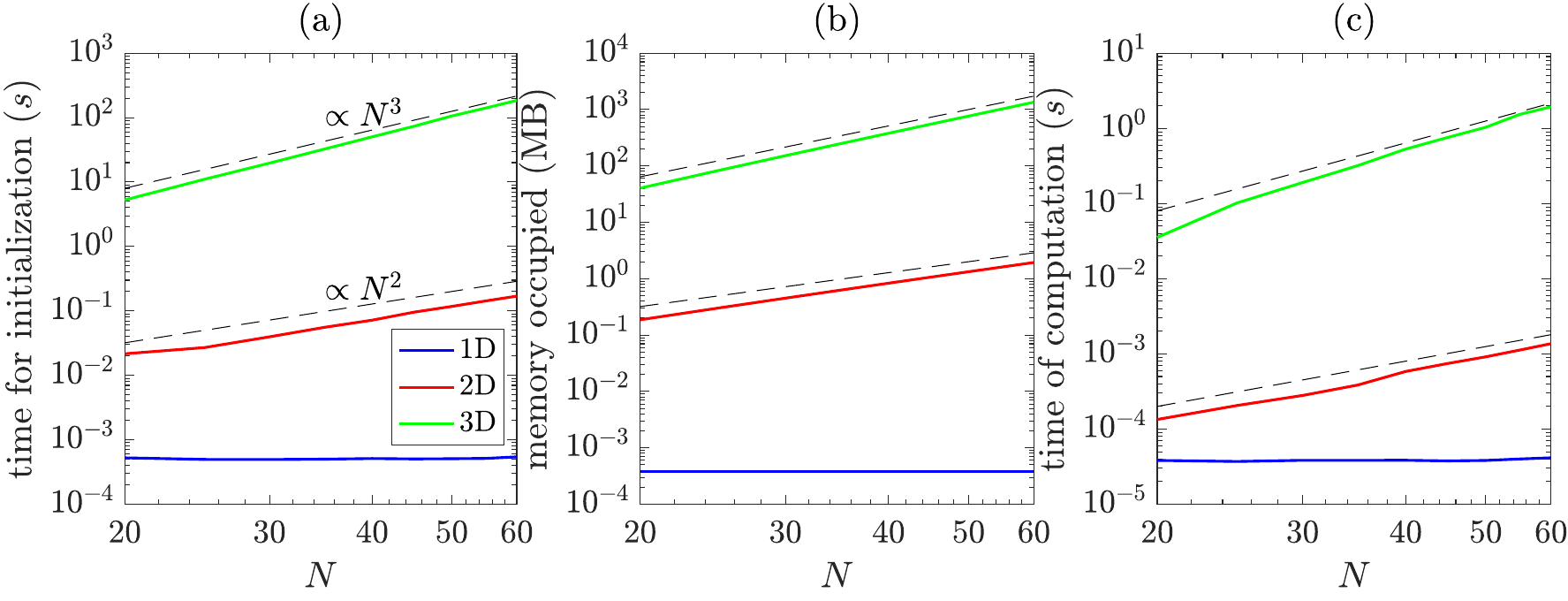}
		\caption{Computational cost of CPU time, in seconds, and memory usage, in MegaBytes, of the product routines with respect to the number of node points $N$, for one-, two- and three-dimensional lattices with golden mean spacing~\ref{L:golden}: (a)~time for initialization of product routines; (b)~memory occupied by the product function handle; (c)~time of computation for one product~\eqref{EQ:convolution}. Dashed lines indicating growths $\propto N^d$ for $d=2,3$ are plotted for comparison. All figures are in log-scale.}
		\label{FIG:computational_cost}
	\end{figure*}

	Fig.~\ref{FIG:computational_cost} summarizes the computational cost of library \textsc{LogLatt} and how this cost grows with the number of points $N$, in one-, two- and three-dimensional lattices; here, the spacing $\lambda$ is the golden mean~\ref{L:golden}.
	Efficiency is estimated up to $N = 60$, which in Fourier space covers a spatial range $k_{\max} = \varphi^{60} \approx 10^{12}$.
	The CPU time of execution is measured in seconds and all runs were performed in \textsc{Matlab} R2016b on a \textsc{Mac}\textsuperscript{\circledR} with Intel\textsuperscript{\circledR} Core i5 CPU $1.8$ GHz $8$ GB RAM.
	
	Figs.~\ref{FIG:computational_cost}(a,b) show the computational cost, in CPU time and memory occupied, for the initialization of the products.
	The three-dimensional lattice with best resolution ($N = 60$) takes around a couple of minutes to initialize and occupies $1$ GB in memory.
	In turn, one- and two-dimensional products are initialized within less than a second and take no more than $2$ MB in memory.
	We recall that the products need to be initialized only once before computations and have the alternative of being saved as mat-files and quickly loaded whenever needed.
	The expense of this initialization is rewarded in the reduced CPU time for a single convolution, as shown in Fig.~\ref{FIG:computational_cost}(c), where, even in the higher three-dimensional resolution, the operation takes not even two seconds to be fully executed.
	Differently from a full convolution, the local convolution~\eqref{EQ:convolution} in the $d$-dimensional lattice has complexity $O(N^d)$. This is readily confirmed by the measured time of computation in Fig.~\ref{FIG:computational_cost}(c), except for the one-dimensional case, which is coded using a different strategy as said above and benefits from \textsc{Matlab}'s multithreading.
	
	\section{Applications to fluid dynamics}\label{SEC:fluid}
	
	The library~\textsc{LogLatt} has been already applied to several important problems in fluid dynamics: blowup and shock solutions in the one-dimensional Burgers equation~\cite{campolina2019fluid}, the chaotic blowup scenario in the three-dimensional incompressible Euler equations~\cite{campolina2018chaotic,campolina2021fluid}, and turbulence in the three-dimensional incompressible Navier-Stokes equations~\cite{campolina2021fluid}.
	A possible extension to isentropic compressible flow~\cite{campolina2021fluid} was also considered.
	Here we show how to implement the library applicabilities on two classical equations, in order to validate the library and attest its efficiency.
	
	In Section~\ref{SEC:Burgers} we apply our library to the one-dimensional Burgers equation and recover some classical results for the dyadic shell model.
	In Section~\ref{SEC:Euler} we sketch an example of application to a more complicated multi-dimensional system by considereing the two-dimensional incompressible Euler equations.
	Here, we present post-processing of the solution and verify conservation laws.
	
	\subsection{One-dimensional Burgers equation}\label{SEC:Burgers}
	
	The forced Burgers equation on the one-dimensional logarithmic lattice~\eqref{EQ:computational_lattice} is given by
	\begin{equation}
	\partial_t u + u \ast \partial_x u = \nu \partial_x^2 u + f,
	\label{EQ:Burgers}
	\end{equation}
	where $u(k,t)$ represents the velocity modes on the lattice $k \in \mathbb{\Lambda}$ at time $t \in \mathbb{R}$, $\nu \geq 0$ is the viscosity and $f(k,t)$ is an external force.
	All functions are supposed to satisfy the reality condition~\eqref{EQ:reality}.
	
	Model~\eqref{EQ:Burgers} on the lattice retains many properties of the original model, like the symmetries of time and space translations $t \mapsto t + \tau$ and $u(k) \mapsto e^{-ik\xi}u(k)$ for any $\tau,\xi \in \mathbb{R}$.
	Inviscid ($\nu = 0$) and unforced ($f = 0$) regular solutions also conserve the energy $E(t) = \frac{1}{2}\Vert u \Vert^2$, for the $\ell^2$ norm~\eqref{EQ:l2norm}, and the third-order moment $H(t) = (u \ast u, u)$, which is well-defined because of associativity in average of the product~\ref{DEF:PROD_associativity_avg} and may be related to a Hamiltonian structure in some shell models~\cite{l1999hamiltonian}.
	All of these properties can be proved using the operations on the lattice -- see~\cite{campolina2019fluid} for details.
	
\begin{lstlisting}[float=b, style=Matlab-bw, basicstyle=\fontsize{10}{10}\mlttfamily, mlcommentstyle= {\color[RGB]{128,128,128}}, caption={m-file Burgers1D.m},captionpos=b, linerange={1-10}, frame=tb, numbers=none, label=MAT:Burgers]
% Burgers1D.m   Solver for Burgers' equation on the one-dimensional logarithmic lattice
N = 20; % number of lattice nodes
[product,l2norm,l2inner,sup,dx,lapl,K] = LogLatt1D(N,'dyadic'); % preamble

nu = 1e-2; % viscosity
f = zeros(N,1); f(1) = 1j; % constant in time force
dudt = @(t,u) -product(u,dx(u)) + nu*lapl(u) + f; % Burgers' equation
uinit = zeros(N,1); % initial condition
T = 5; % time of integration
[t,u] = ode15s(dudt,[0 T],uinit); % solver
\end{lstlisting}
	
	When considering purely imaginary solutions, the Burgers equation~\eqref{EQ:Burgers} on the dyadic lattice~\ref{L:dyadic} can be reduced to the Desnyansky-Novikov shell model of turbulence~\cite{desnyansky1974evolution}, also called the dyadic shell model -- consult~\cite{campolina2021fluid} for the detailed deduction.
	Introducing a constant-in-time force
	\begin{equation}
	f = i\delta_1,
	\label{EQ:force}
	\end{equation}
	where $\delta_1(1) = 1$ and $\delta_1(k) = 0$ for $k \neq 1$, this model is well-known for blowing up when $\nu = 0$ and to recover assymptotically the fixed-point solution $u(k) = ik^{-1/3}$ in the inviscid regularization $\nu \to 0$, a behavior which was related to the development of shock solutions in the original Burgers equation~\cite{mailybaev2015continuous}.
	
	The m-file {\mlttfamily Burgers1D.m} in Listing~\ref{MAT:Burgers} solves the Burgers equation~\eqref{EQ:Burgers}, with the constant forcing~\eqref{EQ:force} and zero initial condition, on the one-dimensional dyadic logarithmic lattice~\ref{L:dyadic}.
	We fix $N = 20$ points in the lattice, which covers a spatial range $k_{\max} = 2^{20} \approx 10^6$.
	The functionalities of the library are initialized through {\mlttfamily LogLatt1D.m} in the preamble, which is called as in~\ref{I_str}.
	The viscosity {\mlttfamily nu} and the force {\mlttfamily f} are defined in the usual way.
	Using the {\mlttfamily product} and {\mlttfamily dx} function handles, the time variation in the Burgers equation is written, as a function of the time {\mlttfamily t} and the velocity {\mlttfamily u}, in the very intuitive way {\mlttfamily dudt = @(t,u) $-$product(u,dx(u))$+$nu*lapl(u)$+$f}.
	For simplicity, we use the \textsc{Matlab} native ODE solver {\mlttfamily ode15s}, a variable-step, variable-order solver for stiff equations based on the numerical differentiation formulas of orders $1$ to $5$ -- consult~\cite{shampine1997matlab} for details.
	{\mlttfamily Burgers1D.m} returns the solution {\mlttfamily u} at time instants {\mlttfamily t} after around $0.06$ second of CPU time.
	
\begin{figure*}[t]
		\centering
		\includegraphics[width=\textwidth]{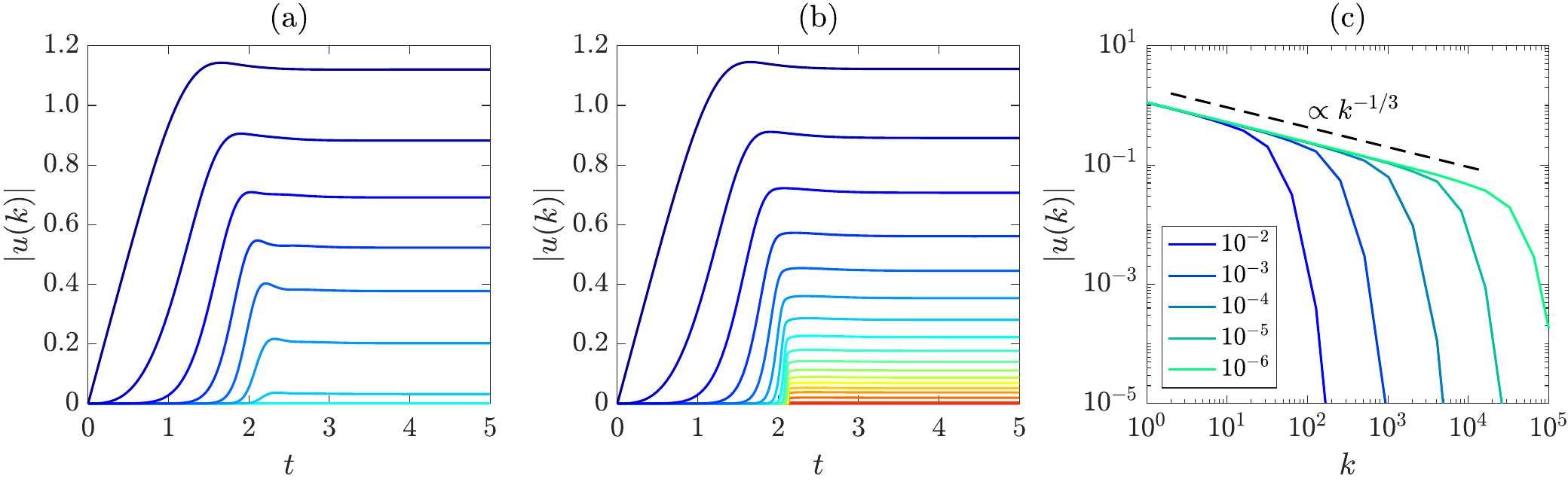}
		\caption{Solutions of the Burgers equation on the one-dimensional dyadic logarithmic lattice~\ref{L:dyadic}, with the constant forcing~\eqref{EQ:force} and zero intial condition: (a)~and (b) show the time evolution of lattice variables $|u(k)|$, at several points $k$, for viscosities $\nu = 10^{-2}$ and $10^{-6}$, respectively; colors change from blue to red by increasing $|k|$; (c)~solution spectrum $|u(k)|$, in log-scales, at the final instant $t = 5$ for different viscosities $\nu = 10^{-2}, \ 10^{-3}, \ 10^{-4}, \ 10^{-5}, \ 10^{-6}$.}
		\label{FIG:burgers}
	\end{figure*}

	Figs.~\ref{FIG:burgers}(a,b) show the time evolution of velocities $|u(k)|$ at several lattice points $k$ for viscosities $\nu = 10^{-2}$ and $10^{-6}$, respectively.
	These graphs are generated by the command {\mlttfamily plot(t,abs(u))}.
	We observe an abrupt growth of the velocities around the inviscid blowup time $t_b \approx 2.13$, which gets more pronounced for smaller viscosities.
	The viscous solution can be extended beyond the blowup time and develops an assymptotic power-law scaling $|u(k)| \sim k^{-1/3}$ as the dissipation range is shifted towards larger $k$ in the inviscid limit $\nu \to 0$.
	This dynamics is better visualized by plotting, in log-scales, the solution spectrum at the final instant of time through the command {\mlttfamily loglog(K,abs(u(end,:)))} and is readily verified in Fig.~\ref{FIG:burgers}(c).
	
	\subsection{Two-dimensional incompressible Euler equations}\label{SEC:Euler}
	
\begin{lstlisting}[float=t, style=Matlab-bw, basicstyle=\fontsize{10}{10}\mlttfamily, mlcommentstyle= {\color[RGB]{128,128,128}}, caption={m-file Euler2D.m.},captionpos=b, linerange={1-26}, frame=tb, numbers=none, label=MAT:Euler]
% Euler2D.m  Solver for the incomp. Euler equations on a two-dimensional logarithmic lattice
N = 20; % number of points in each direction
product = LogLatt2D_product(N,'golden'); % product on the logarithmic lattice
[dx,dy,lapl,lapl_,grad,div,rot,rot_] = LogLatt2D_diff(N,'golden'); % differential operators

array2mat = @(w) reshape(w,[N,N,2]);   % adjust data from array  to matrix
mat2array = @(w) reshape(w,[2*N^2,1]); % adjust data from matrix to array

dwdt = @(t,w) RHS(w,product,dx,dy,rot_,array2mat,mat2array); % Euler equations
winit = zeros(N,N,2); % initial condition
randn('state',100); % set the state of randn
winit(2:4,2:4,:) = randn(3,3,2) + 1j*randn(3,3,2); % random initial condition
winit = mat2array(winit); % adjust data of initial condition
T = 10; % time of integration
[t,w] = ode15s(dwdt,[0 T],winit); % solver
w = reshape(w,[length(t),N,N,2]); % adjust array size

% RHS.m Right-hand-side of the two-dimensional incomp. Euler equations
function dwdt = RHS(w,product,dx,dy,rot_,array2mat,mat2array)
    w = array2mat(w); % adjust data from array to matrix
    u = rot_(w); % compute velocities from vorticity
    ux = u(:,:,:,1); % x velocity
    uy = u(:,:,:,2); % y velocity
    dwdt = -product(ux,dx(w)) - product(uy,dy(w)); % Euler equations
    dwdt = mat2array(dwdt); % adjust data from matrix to array
end
\end{lstlisting}
	
	We present now the application of \textsc{LogLatt} to the incompressible Euler equations on the two-dimensional logarithmic lattice $\mathbb{\Lambda}^2$.
	As usual in Direct Numerical Simulations (DNS), we solve the Euler equations on the lattice in vorticity formulation
	\begin{equation}
	\partial_t \omega + u_x \ast \partial_x \omega + u_y \ast \partial_y \omega = 0,
	\label{EQ:Euler}
	\end{equation}
	where $\omega(\mathbf{k},t) \in \mathbb{C}$ represents the scalar vorticity and $\mathbf{u}(\mathbf{k},t) = (u_x,u_y) \in \mathbb{C}^2$ the incompressible velocity field on the lattice $\mathbf{k} \in \mathbb{\Lambda}^2$, both satisfying the reality condition~\eqref{EQ:reality}.
	The velocity is computed from the vorticity by the two-dimensional Biot-Savart law on the lattice
	\begin{equation}
	\mathbf{u} = \text{rot}^{-1} \ \omega = -\Delta^{-1}(\partial_y \omega, -\partial_x \omega).
	\label{EQ:Biot-Savart}
	\end{equation}
	Observe that $\mathbf{u}$ obtained from formula~\eqref{EQ:Biot-Savart} is always a solenoidal vector field, \textit{i.e.}, $\text{div} \ \mathbf{u} = 0$.
	
	The Euler equations on the lattice keep many properties of the original model. These include not only incompressibility and a similar group of symmetries, but also finer properties of ideal flow, like Kelvin's Theorem~\cite{campolina2021fluid} and, in the three-dimensional case, correlation of solutions with DNS results~\cite{campolina2018chaotic}.
	Particularly, the two-dimensional model~\eqref{EQ:Euler}-\eqref{EQ:Biot-Savart} conserves in time the energy
	\begin{equation}
	E(t) = \frac{1}{2}\Vert\mathbf{u}\Vert^2,
	\label{EQ:Energy}
	\end{equation}
	and the enstrophy
	\begin{equation}
	\Omega(t) = \frac{1}{2}\Vert\omega\Vert^2,
	\label{EQ:enstrophy}
	\end{equation}
	where $\Vert\cdot\Vert$ is the $\ell^2$ norm~\eqref{EQ:l2norm}.
	
	\begin{figure*}[t]
		\centering
		\includegraphics[width=.65\textwidth]{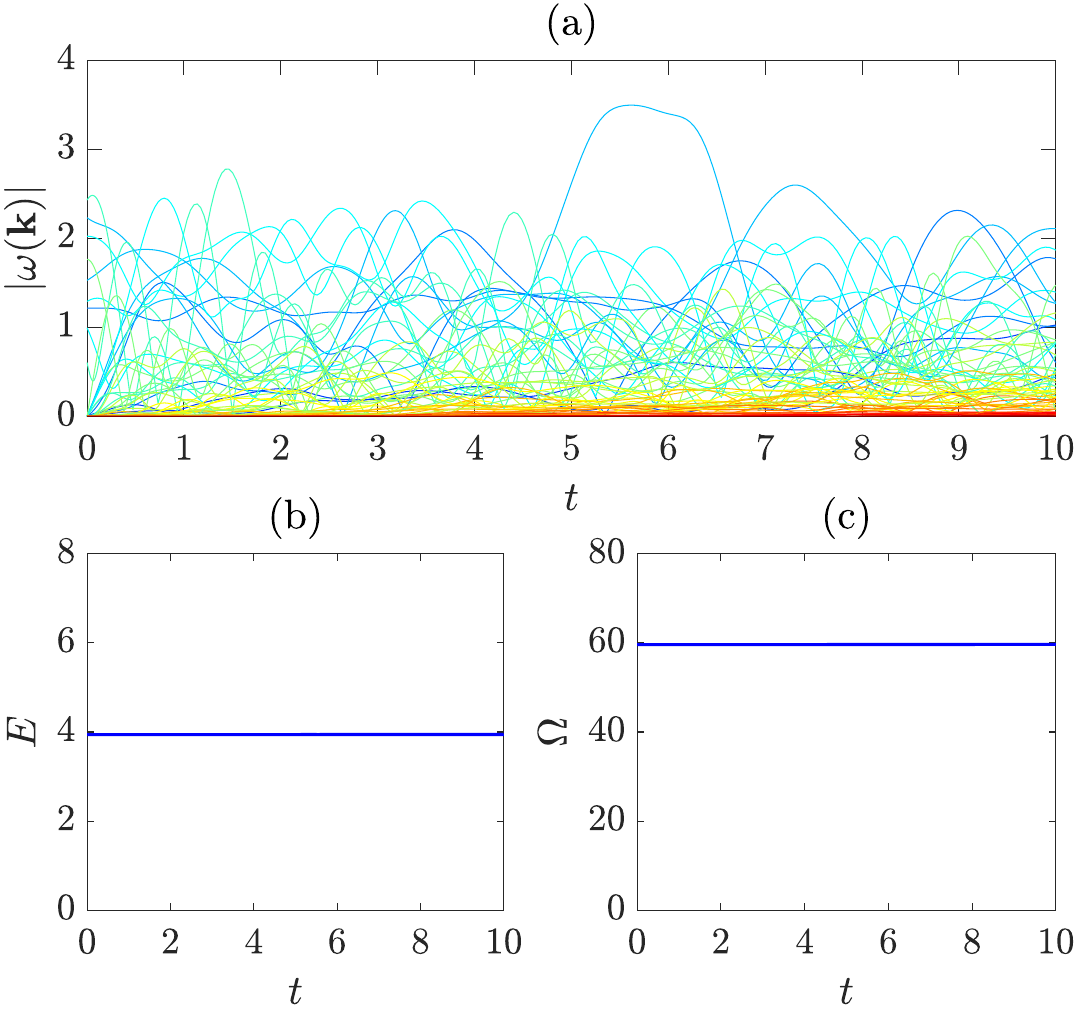}
		\caption{Solution of the incompressible Euler equations on the two-dimensional logarithmic lattice with golden mean spacing~\ref{L:golden}: (a)~time evolution of scalar vorticities $|\omega(\mathbf{k})|$ at different points $\mathbf{k}$; colors change from blue to red by increasing $|\mathbf{k}|$; (b)~time evolution of the energy~\eqref{EQ:Energy}; (c)~time evolution of the enstrophy~\eqref{EQ:enstrophy}.}
		\label{FIG:Euler}
	\end{figure*}
	
	The m-file {\mlttfamily Euler2D.m} in Listing~\ref{MAT:Euler} solves the incompressible Euler equations in vorticity formulation on the two-dimensional lattice of golden mean spacing~\ref{L:golden}.
	We set the number of points $N = 20$ in each direction, which provides a spatial range of $k_{\max} = \varphi^{20} \approx 10^4$.
	The \textsc{LogLatt} applicabilities used in this routine are the product and some differential operators, initialized in the preamble from {\mlttfamily LogLatt2D\_product.m} and {\mlttfamily LogLatt2D\_diff.m}, both called as in~\ref{I_str}.
	Here, the Euler equations are encoded in the nested function {\mlttfamily RHS}.
	The velocity components {\mlttfamily ux} and {\mlttfamily uy} are computed from the scalar vorticity {\mlttfamily w} by the Biot-Savart law~\eqref{EQ:Biot-Savart} through the inverse rotational operator {\mlttfamily rot\_}.
	Vorticity equation~\eqref{EQ:Euler} is written in the very clear way {\mlttfamily dwdt = $-$product(ux,dx(w))$-$product(uy,dy(w))}.
	We employ again the {\mlttfamily ode15s} solver, which operates on column arrays only.
	The functions {\mlttfamily array2mat} and {\mlttfamily mat2array} defined on the preamble are responsible for reshaping column arrays to matrices and vice-versa.
	We initialize the flow with random complex vorticities at large scales.
	In order to make experiments repeatable, we fix the initial state of the random number generator with the command {\mlttfamily randn(\textquotesingle state\textquotesingle,100)}.
	{\mlttfamily Euler2D.m} runs within approximately $6$ seconds.
	
	Fig.~\ref{FIG:Euler}(a) shows the time evolution of vorticities $|\omega(\mathbf{k})|$ at different lattice points $\mathbf{k}$.
	The chaotic Euler dynamics of individual lattice variables contrasts with the ordered layered behavior of Burgers velocities in Fig.~\ref{FIG:burgers}(a,b).
	Using the {\mlttfamily l2norm} function handle output from {\mlttfamily LogLatt2D\_norms.m}, we compute the energy~\eqref{EQ:Energy} and the enstrophy~\eqref{EQ:enstrophy} at each time step with the sample code in Listing~\ref{MAT:post}.
	
\begin{lstlisting}[float=t, style=Matlab-bw, basicstyle=\fontsize{10}{10}\mlttfamily, mlcommentstyle= {\color[RGB]{128,128,128}}, linerange={1-10}, caption={A sample code for the computation of energy and enstrophy from the solution w at t from Euler2D.m.},captionpos=b, frame=tb, numbers=none, label=MAT:post]
[l2norm,l2inner,sup] = LogLatt2D_norms; % norms
E = zeros(size(t)); % energy
Z = zeros(size(t)); % enstrophy
for jj = 1:length(t)
    ww = w(jj,:,:,:);         % vorticity at step jj
    ww = reshape(ww,[N,N,2]); % remove residual dimension of length 1
    u = rot_(ww);             % velocity at step jj
    E(jj) = .5*l2norm(u)^2;   % energy computation
    Z(jj) = .5*l2norm(ww)^2;  % enstrophy computation
end
\end{lstlisting}
	
	Figs.~\ref{FIG:Euler}(b,c) display constant values of energy and enstrophy along time, and thus attest the conservation laws for the Euler equations.
	
	\section{Conclusions}\label{SEC:conclusions}
	
	We introduced \textsc{LogLatt}, a \textsc{Matlab} library for the computational calculus on logarithmic lattices.
	The applicabilities are split into several modules for one, two and three dimensions and comprise common differential operators from vector calculus, norms and the functional product, which is a local convolution on the lattice.
	These functionalities are available for all possible lattice spacings classified in~\cite{campolina2021fluid} and are encoded as \textsc{Matlab} function handles, so their implementations result in elegant and intuitive scripts.
	The operations are accurately executed within a small time of computation.
	We validated the library by recovering classical dynamical results of the Burgers equation on the dyadic logarithmic lattice and through the verification of some conservation laws in the two-dimensional incompressible Euler equations.
	Using a few simple lines of code, \textsc{LogLatt} solves these equations within seconds of computation, and their solutions still cover a large spatial range, possibly inaccessible to the current direct numerical simulations.
	\textsc{LogLatt} is freely available for noncommercial use in \textsc{Matlab} Central File Exchange~\cite{campolina2020loglattmatlab} and is readily applicable in the study of partial any differential equation with quadratic nonlinearity.
	
	\section*{Acknowledgment}
	
	The author thanks A. Mailybaev for useful suggestions and for following closely the development of this library.
	The many improvements on the text by E. Trist\~{a}o are also greatly acknowledged.
	The author dedicates this work to the Blessed Virgin Mary.
	
	\bibliographystyle{plain}
	\bibliography{refs}
\end{document}